# The Construction and Globalization of the Knowledge Base in Inter-human Communication Systems

Loet Leydesdorff
Science & Technology Dynamics, University of Amsterdam
Amsterdam School of Communications Research (ASCoR)
Kloveniersburgwal 48, 1012 CX  Amsterdam, The Netherlands
loet@leydesdorff.net; http://www.leydesdorff.net/

**Abstract**

The relationship between the "knowledge base" and the "globalization" of communication systems is discussed from the perspective of communication theory. I argue that inter-human communication takes place at two levels. At the first level information is exchanged and provided with meaning and at the second level meaning can reflexively be communicated. Human language can be considered as the evolutionary achievement which enables us to use these two channels of communication simultaneously. Providing meaning from the perspective of hindsight is a recursive operation: a meaning that makes a difference can be considered as knowledge. If the production of knowledge is socially organized, the perspective of hindsight can further be codified. This adds globalization to the historically stabilized patterns of communications. Globalization can be expected to transform the communications in an evolutionary mode. However, the self-organization of a knowledge-based society remains an expectation with the status of a hypothesis.

## Introduction

A body of literature is currently devoted to the globalization of the economy and the so-called knowledge-based society (e.g., David & Foray, 2002; Leydesdorff, 2001; Mokyr, 2002). How can the concepts of "knowledge based" and "globalization" be theoretically related? In this paper, I argue that communication theory is needed for understanding this relationship. A knowledge base is generated within (for example, scientific) communication when another, that is, global, dimension can be added to the communication. Globalization of the communication operates selectively upon stabilizations that are shaped historically.

Language can be considered as an evolutionary achievement that enables us to construct meaning at the supra-individual level. First, the expected information contained in a message can be provided with meaning by selecting a "signal" from the "noise." Once meaning has been attributed to the signal locally, this can again be communicated. Recursively some meanings can be selected as more meaningful than others by using codifications of meaning in the communication. A meaning that makes a difference with reference to a code can be considered as knowledge. In other words, the construction of discursive knowledge is a result of using language in increasingly codified communication.

Meaning is provided reflexively, that is, with hindsight. The emergent order of expectations is constructed and continuously reconstructed on the basis of the communications that have been organized previously. Although a knowledge-based order remains emergent and historically embedded, the codification in knowledge-based systems of communication provides a basis for the reconstruction of the social system different from its historical construction. The subdynamics of historical construction and reflexive reconstruction operate with different perspectives on time (Urry, 2000).

As knowledge production became increasingly organized in scientific discourses during the past two centuries (e.g., Gilbert & Mulkay, 1984; Leydesdorff, 1995; Stichweh, 1984), the knowledge-based perspective has become an institutionalized subdynamic of the social system (Whitley, 1984). From this perspective, the historical configurations provide the communication systems with stabilities needed for the meta-stable or globalized competition in a next-order layer. This layer, however, can only be accessed through a knowledge-based representation.

In order to make my argument that the evolutionary reorganization of communication can increasingly operate as a feedback on the historical development of the social system, I shall proceed step-by-step. Following Luhmann (1984), I first distinguish between action systems (Parsons, 1937) and communication systems as different types of systems. Communication systems process information differently from action systems and therefore can be expected to provide information with another kind of meaning. In a second step, I distinguish between observed and expected information. By elaborating Shannon's (1948) mathematical theory of communication into a non-linear dynamics of communication, a more abstract concept of meaning can be defined. The conditions for developing a knowledge-based system of communication can then be specified.

**First- and second-order observations**

Whereas Parsons's (1937) theory of social systems focused on actions at the nodes of the network, Luhmann's (1984) sociological theory of communication, in contrast, focuses on the links that potentially coordinate the network into an evolving system. The links can be considered as the operators of a network that emerges as a system among the nodes (Barabási, 2002). The links provide momentary structure to the network. In social network analysis, for example, this structure is examined in terms of dimensions using factor analysis.

Observable actions can be attributed to agencies. The latter have been considered as units of analysis in the sociological research design. In communication theory, however, one focuses on the communications as operations. Units of operation cannot be observed without a further level of reflection. The communicative operations, for example, may also change the units of analysis. The observables, however, can be analyzed as the traces of previous communications when the communicative operations are properly specified. For example, an institution may have been shaped as the result of a social conflict.

Luhmann (1984, p. 226 [1995, p. 164]) emphasized that *"communication cannot be observed directly, only inferred"* (italics in the original).[1] Some authors have called this reflexive inference "second-order" observations (Bäcker, 1999). Because one needs theoretical guidance for making an inference, a "second-order observation" can be considered as the specification of an expectation. However expectations are, in important respects, different from observations. For example, expectations operate in terms of uncertainties, while observations serve the observer mainly for reducing the uncertainty.

---

[1] Luhmann (1984: 226): "Die wichtigste Konsequenz dieser Analyse ist: daß Kommunikation nicht direct beobachtet, sonder nur erschlossen werden kann."

Consequently, a network system of communications cannot be expected to observe like a human being. The use of anthropomorphic metaphors (e.g., "observer") may easily lead to misunderstandings: not the human observers, but the observations and the observational reports play a key role in the communication. The system of reference is different. When an observation is reported, the observation can contribute to updating the expectations entertained at the network level. As noted, this second-order exchange cannot be analyzed without a theoretical perspective. Observations can then be interpreted discursively in relation to one another. Expectations--"second-order" or theoretically informed observations--can thus be further specified.

If we assume that expectations operate as the substance of communication at the level of social systems, how can such a system be reproduced? In order to answer the question in terms of a model it is first necessary to abstract from specific expectations. In my opinion, Shannon's (1948) mathematical theory of communication can be useful for answering this question because the concept of "expectation" is fundamental to the abstraction in this theory. In Shannon's theory, a distribution (e.g., of links and nodes in a network system) contains an uncertainty that can be communicated as the *expected* information of the message that this configuration has occurred.

The definitions at this level are formal; that is, they do not refer to specific systems of reference. Specification of the systems of reference requires substantive theorizing, for example, about the role of expectations in inter-human communication. In other words, we have first to specify how meaning is generated in communication and then again communicated and reproduced.

**Information, uncertainty, and meaning**

Although sociology and the mathematical theory of communication both focus on distributed systems (e.g., networks), words like "information," "uncertainty," or "communication" mean different things in other intellectual traditions. Furthermore, the distributions in social systems develop at various levels at the same time. The sociological research design therefore needs methodological guidance. However, the mathematical theory of communication has been discredited in the social sciences with the argument that Shannon (1948) defined "information" abstractly as uncertainty (e.g., Bailey, 1994). This definition appears counter-intuitive, since one tends to associate "information" with a message that informs a system receiving the signal. From the perspective of a receiving system, information can be defined only as meaningful information or "a difference that makes a difference" (Bateson, 1972, p. 453; Luhmann, 1984, p. 103).

The use of two concepts for information has resulted in a lot of semantic confusion. Let us first specify the difference between information as uncertainty and information that is meaningful to a system receiving this information. Shannon detached himself from the implications of his definition of information as uncertainty by stating that the "semantic aspects of communication are irrelevant to the engineering problem" (Shannon & Weaver, 1949, p. 3). The Shannon-type information precedes the information formed within a system.

The system of reference is not yet specified (Theil, 1972). The message only informs us that an event (or a series of events) has happened. However, the unit of measurement of the uncertainty thus generated (i.e., bits of information) remains formal and therefore dimensionless.[2]

When a system receives this (Shannon-type) information, it may be disturbed and therefore initially become more uncertain (about its environment for example). By processing the uncertainty internally, the receiving system can sometimes--that is, if it contains substantive dimensions--discard part of the information as noise. The remainder is then selectively designated as meaningful information. After the de-selection of the noise, the meaningful information potentially reduces the uncertainty. Brillouin (1962) proposed to call this decrease in the uncertainty "negentropy." The meaningful information contains a selection on the uncertainty that prevails.

Despite the confusion of the two concepts of information in the literature (Hayles, 1990), the biological theory of autopoietic systems (e.g., Maturana & Varela, 1984) and the information-theoretical approach (e.g., Abramson, 1963; Theil, 1972) have been consistent in excluding each other's definitions of "information" for analytical reasons (Boshouwers, 1997). Biological systems can be considered as "natural," and therefore the biologist is inclined to begin with the specification of an observation rather than the uncertainty of an expectation. As Maturana & Varela (1980, p. 90) formulated it:

> Notions such as coding and transmission of information do not enter in the realization of a concrete autopoietic system because they do not refer to actual processes in it.

While these authors insisted on the biological realization of "actual processes," Shannon's co-author Weaver (1949, pp. 116f.) noted the problem of defining "meaning" from a mathematical perspective when he formulated the following:

> The concept of information developed in this theory at first seems disappointing and bizarre--disappointing because it has nothing to do with the meaning, and bizarre because it deals not with a single message but rather with the statistical character of a whole ensemble of messages, bizarre also because in these statistical terms the two words *information* and *uncertainty* find themselves to be partners.
>     I think, however, that these should be only temporary reactions; and that one should say, at the end, that this analysis has so penetratingly cleared the air that one is now, perhaps for the first time, ready for a real theory of meaning.

My purpose is to abstract from observers as biological systems and to discuss systems of communication that are able to communicate about meaning and expectations in addition to observations. I therefore follow the intuition of Weaver and try to define "meaning" first without reference to any specific "realization."

---

[2] Shannon (1948) defined the uncertainty or information as equal to the probabilistic entropy of the communication system. Thermodynamic entropy, however, is different from probabilistic entropy because the former is measured in terms of Joule/Kelvin, while bits of information are dimensionless.

The meaning of information can be defined only with reference to a system that is able to organize the information. The generation of meaning therefore assumes a system operating over time. The *specification of a system of reference* provides the (Shannon-type) information with system-specific meaning. That is, a system contains a substance in which information is communicated. This substance can also be considered as a medium of the communication. A probabilistic entropy is generated whenever the communication system operates and the medium is consequently redistributed. This Shannon-type information can be measured, but the measurement results could additionally be provided with a substantive interpretation if the system of reference were to be specified.

For example, information theory can be elaborated into a statistics for the study of economic transaction processes (Theil, 1972) or for the study of biological evolution processes (Brooks and Wiley, 1986). Substantive theorizing is needed for the specification of the relevant system(s) of reference (MacKay, 1956) so that another theory of communication --such as one concerning economic exchange relations--can be generated (Steinmueller, 2002). The mathematical theory of communication can be used as a formal methodology for studying the non-linear dynamics of substantive systems thus specified.

**Luhmann's sociological theory of communication**

Luhmann (1984 and 1986) distinguished between individual (psychological) systems and social systems. Both social and psychological systems were specified as substantively different from biological systems to the extent that they are able to process expectations by providing observations with meaning. However, when they receive signals, these two types of systems can be expected to use different dynamics for their respective update. This is because the concept of "observation" has a different status in these two contexts. The social system can only process information and meaning in a distributed mode; that is, by communicating it as the *expected* information content of the communication.

This exchange of the expectations provided us above with the second-order domain. This domain is empirical, but reflexive. That is to say the observation at this level remains an expectation that can only be made by providing a theoretical interpretation to first-order observations. Thus, the second-order domain emerges within the first-order and it remains referential to it. One is thereby able to observe not only what the observers under study are observing, but also how these observers are providing meaning to their observations and communicating about this. Furthermore, when observations are attributed as one type or the other, this attribution can be expected to remain uncertain. The attributions can be de-constructed and reconstructed through further communication.

The opportunity to distinguish between the exchange of expectations and the reports of observations enables us to redefine the relationship between autopoietic systems theory and information theory at the level of social systems. The exchange relations take place at two levels or, in other words, in different dimensions of the communication. These dimensions may additionally interact. Giddens (1979) has called this duality a double hermeneutics, but he juxtaposed the two hermeneutics in terms of participant-observers and external observers. When one appreciates the two levels as dimensions of the communication, the interaction terms can also be specified.

In a process that communicates both meaning and information, uncertainty is generated in more than a single dimension. Using information theory, the uncertainties in different dimensions can be measured as multi-variate probability distributions. As noted, understanding the meaning of the results requires that the system(s) of reference be specified properly. This theoretical specification has to be complex enough to account for the information contained in the messages in the various dimensions. If there are two dimensions, one can analyze both of them *and* their co-variation.

It is important to note that the specification of uncertainty in terms of a (second-order) system of expectations may become abstract, because a discourse about expectations no longer refers exclusively to concrete (first-order) observables, but also to meanings attributed to these observables. Furthermore, a social distribution of expectations can be provided with meaning recursively. Recognizing meaning as one among any number of possible meanings opens a horizon of expectations (Husserl, 1939; Luhmann, 1984, pp. 114ff.). I shall argue below that this next-order relationship of meaning to other possible meanings can be distinguished from the historical generation of meaning.

**Stabilization and globalization**

The provision of meaning requires an operation over time by a receiving system, but this system does not yet have to be stable for any period longer than this single operation. A communication system may also be volatile. In order to be identified as a system, it is assumed that the system under study updates in terms of *repeatedly* distinguishing between signals and noise. In this model, a time axis representing selections diachronically is declared that stands perpendicular to the selections made by the system at each moment in time (Figure 1).

Meaning reduces the uncertainty by selecting from incoming information over time. The variation over time (or "change") can be considered as a diachronic outcome of the interactions between variation and selection at different moments in time. Selection is a recursive operation. Meaning processing systems can be expected to select some of the previous selections for stabilization. Since selections reduce the uncertainty; stabilization reduces the uncertainty with a next order of magnitude. At a next round, stabilizations can further be selected for globalization.

Potentially different stabilizations can again be considered as observable variations, but this would imply the perspective of a next-order selecting system. That is, a distinction between variation and stabilization is made. The next-order, or globalized system, is able to select among the results of the first-order stabilizations. Globalization can thus be considered as a second-order selection process. The next-order system builds upon the lower-level ones by selecting among them and by potentially rewriting the previously attributed meanings in terms of their relative weights in the distribution of possible meanings.

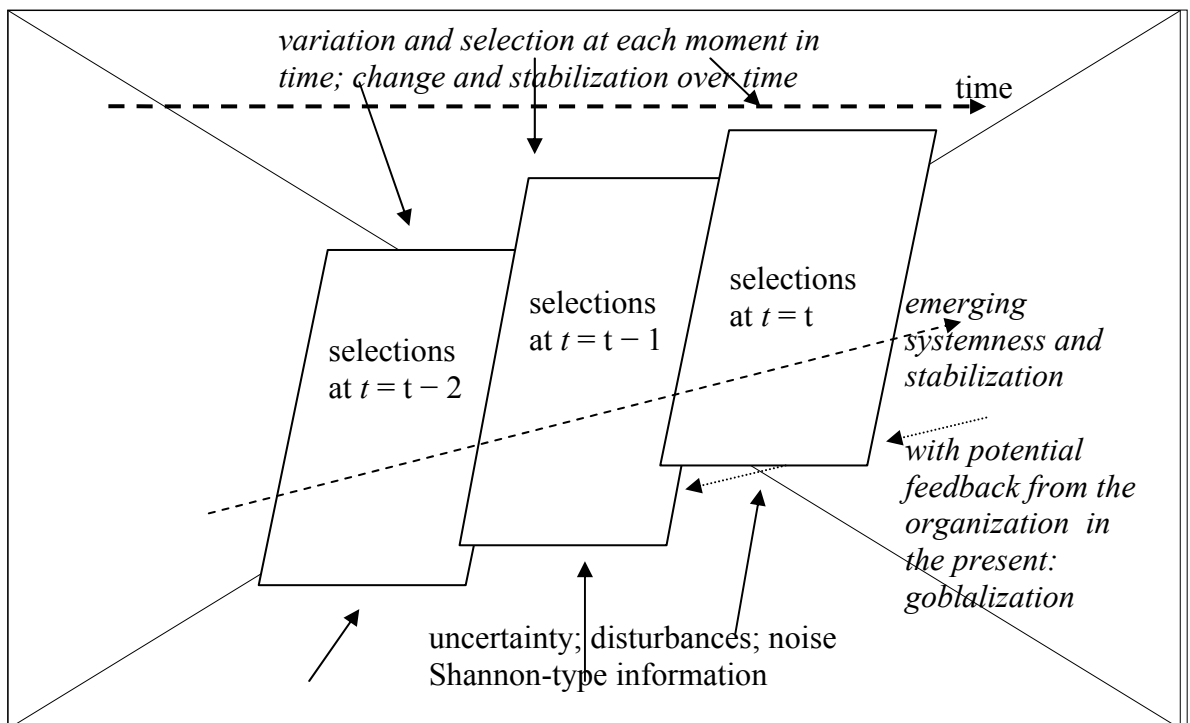

**Figure 1**
*Selection* of variation at each moment in time; *stabilization* of change over time; *globalization* as a feedback mechanism in the present

The recursivity of the operation of providing meaning by a next-order system adds another degree of freedom to the system. However, this globalization of the system is not completed, but remains under construction on top of the previous stabilizations. Stabilization and globalization can be considered as different subdynamics operating on each other. The constructing system comes under the selection pressure of its global dimension as specific constructions are historically realized among the other possible constructions. As a system is shaped historically along the time dimension in a forward mode, global meaning can be attributed only with hindsight. This is achieved by comparing the historical system with other envisioned possible systems. This operation is knowledge-based because the alternatives were not yet all realized. However, the next-order selection operates upon their temporarily stabilized representations in the present.

A reflection operating with reference to an instance invoked in the present ($x_t$) can be distinguished from the recursive update of meaning operating with reference to a historically previous state of the system ($x_{(t-1)}$). Dubois (1998) proposed to call this recursion on a present state of the system an "incursion." An incursion evaluates a representation from a next-order perspective. However, the operation remains historical in the evolution of the system, and an incursive system can therefore be expected to develop recursively along the time axis. "Incursion" occurs *within* systems under historical conditions; that is, as an empirical relation to recursively constructed and reconstructed trajectories.

An incursive system is able to select among its current representations of the past in terms of its future "survival value" in a next-order selection environment (Rosen, 1985). Incursion and recursion can be expected to intertwine as modes of communication in a social system that operates in terms of expectations. Analytically, however, incursion provides the social system with a sub-dynamic different from historical recursion since the incursive system operates on a representation of the system in the present. Because of the ability to select among a variety of possible representations of the system, an incursive system can learn to "anticipate" possible further developments and, therefore, become increasingly "self-organizing." Additionally, a meaning-processing system can construct (knowledge-based) representations that compete with those previously generated.

Unlike artificial systems, social systems remain historical and therefore referential to first-order observations. The virtual operation of the global dimension can be expected to interact with other dimensions of the system. Because of this interaction with its history, the meaning-processing system can be expected to fail empirically to self-organize. Here, self-organization remains only a subdynamic of the historical system. The social system can be expected to exhibit a mixture of contingencies of its construction in history and its next-order incursions, upon these manifestations. Analytically, however, a double-layered structure sustaining the exchange of both information and meaning is required before the "globalization" of a system of expectations can become historically important.

At the first level, variation is selected and then selections can be chosen for stabilization so that the structural advancements of the networks can also be retained historically. The stabilizations function evolutionarily as retention mechanisms that have been shown to provide "survival value" for the configurations that were realized. At a second level, these historical configurations can further be selected as second-order variations by another (next-order) dimension of the system. This globalizing operation of the system remains "virtual" from a first-order perspective (Giddens, 1984) because globalization incurs on the observable instantiations of the system using another dimension of that system. The additional dimension is spanned in terms of the distribution of observable instances.

**Codification and the change of meaning**

Whereas positive theories are able to specify observable variation (at each moment in time) and change (over time), the specification of a "selection" requires an analytic perspective. Selection is a negative operation: it reduces the variation. In that a negative operation cannot be directly observed, a theoretical inference is needed. Stabilizations can again be observed. The negative sign of the selection can be expected to lead alternately to "observable" and "expected" events with each consequent turn.

Stabilizations can be distinguished analytically from lower-level variations when the selections that preceded the former are analytically specified. Globalization does not require a single stabilization since a great variety can be selected from. With hindsight one can ask whether an identified unity has remained the same despite ongoing processes of change. However an identifiable system can be expected to develop along an observable trajectory, while under selection pressure from a next-order "regime" (Dosi, 1982).

The next-order regime remains pending as a latent form of selection pressure on the systems and subsystems upon which it rests. By attributing an analytical identity to this next-order system the analyst would reduce its complexity by choosing a perspective. The self-organization of the *social* system cannot be identified empirically because it can be expected to remain uncertain and therefore other appreciations remain possible.[3] Thus, the identification of a global system is analytical: the self-organization of a social system can be specified only as a hypothesis. The question then becomes whether entertaining this hypothesis adds to our theoretical understanding.

In other words, the self-organization of the social system beyond its stabilization, is an intrinsically knowledge-based assumption. What is observable provides us with fragments of the global system, which can only be appreciated as instantiations on the basis of entertaining the hypothesis of other options. In this context, Luhmann (1984) proposed to distinguish among three representations of the social system relevant for pursuing analysis: society as the global system of communication, the historical organization of communication, and concrete interactions.

By using social network analysis, interactions can be analyzed as being organized by latent dimensions (Burt, 1982; Lazarsfeld & Henry, 1968). This organization of the interactions can be appreciated by a reflexive actor. However, the further self-organization of the interactions by a next-order level of global organization assumes another dynamic at the structural level. The flux in the latent dimensions may be experienced as counter-intuitive by informed and reflexive actors. The interpretation can then only be knowledge-based.

Knowledge based, as for example in the case of a child who knows that its parents entertain a level of communication to which it has no access. Here the child may begin to understand that this next-order level communication can be expected to redefine its own situation beyond its control or understanding. Whether the family is stabilized or destabilized by the communication between the parents remains structurally beyond the children's control. However, the situation can be reflected in a therapeutic setting.

A globalized system of communications self-organizes the interactions and organizations subsumed under it by selecting (in the present) using the hypothesized degree of freedom for a next-order reflection. By specifying and entertaining this hypothesis, one is able to distinguish analytically between a system developing its complexity historically, that is, along a trajectory, and a self-organizing system that has one more degree of freedom for adjusting to its environments from the perspective of hindsight. When this additional degree of freedom is used reflexively within the system as a dimension of uncertain communication, a meaning-processing system is generated on top of the information-processing one, on which it reflects.

In a meaning-processing system, meaning can be changed by being communicated. An interaction term is generated that may provide feedback on the definitions. The first-order exchange of information is provided with meaning that can further be exchanged at a next-order layer, and this potentially changes the system. Whereas the Shannon-type information provided us only with distributions and differences, "the difference that makes a difference" can then be communicated as "meaningful information," i.e., an interaction term between meaning processing and information processing.

---

[3] An identity can perhaps be defined as a codified and, therefore, symbolically stabilized system that is able to entertain its relation to its own next-order system reflexively without loosing stability.

**Language and culture**

I submit that language can be considered the evolutionary achievement that enables us to attach meaning to the exchange of information and to exchange meaning at the same time (Luhmann, 1984, p. 208 [1995, p. 152]). The analytical distinction is continuously blurred in observable practices because language enables us also to operate informally and in a network mode. A message can be provided with meaning, and the meaning can be distinguished from the expected information contained in the message. Thus, through the linguistic ability communication processes are doubled into two interacting layers within the exchange. This occurs however, in an uncertain mode.

The differentiation between the two channels of communication--uncertain information and meaning--is not "hard-wired." Inter-human communication can therefore be expected to remain prone to failure. Where codification helps us to keep the various meanings separate, the uncertainty introduced by this "double hermeneutics," allows both for formalization and for informal communication. The two layers can be expected to interact continuously.
In other words, one can always consider meaningful what has been previously communicated as uncertainty, and one can discuss and deconstruct discursively what previously was codified. The exchanges of meaning and uncertainty co-vary in the communication of "meaningful information." The various exchange processes can be expected to generate uncertainty (i.e., introduce new variation) by interacting.

Participants and observers are able to take part both in the historical construction and in the recursive feedback of providing meaning to the reconstructions. This occurs at potentially different times, in different places, and to different degrees. When linguistic capacities enable agents to counteract natural constraints at the supra-individual level by constructing and maintaining a specific codification as a cultural feedback loop, a double hermeneutics is implicated reflexively (Giddens, 1976). The possibility of alternating perspectives between the role of participant-observer and external observer, provides the basis for improving one's analytical understanding of the reflexive operation of interaction (Parsons, 1968). Two (or more) perspectives for the selection can be entertained and traded-off.

A biological system has to adapt--at the risk of extinction--because the selection pressures are "natural." Selection environments, however, can already vary among ecological niches. Biological exchange relations can therefore experience additional selection pressure. Moreover, the two types of selection pressures cannot be distinguished by the agents living in these niches unless these carrying agents have the reflexive capacity to entertain concepts for mapping their situation. When these concepts can be communicated, a semantic domain is generated operationally on top of the consensual domain (in which the agents relate structurally) (Maturana, 1978). The system of communications can then become self-organizing and autopoietic in the sense of increasingly controlling its own reproduction. This happens in biological cultures like insect populations.

Human language extends the biological concept of a semantic domain because order is not constructed in a biological environment and then stabilized, but remains flexible and under construction as an expectation of order. This constructed order can also be changed by a next-order system or at a next moment in time, i.e., by adding reflexively a new dimension to the system. Note that "reproduction" in this case no longer means reproduction in the biological sense, but rather the further development of the communication through the system's previously organized retention mechanisms. The lower-level systems upon which the next-level systems build, can be innovatively reconstructed given the additional degree of freedom provided by linguistic exchanges. The represented and the representing systems may thus begin to co-evolve.

The number of reflexive couplings maintained between the layers, may perhaps be considered as the evolutionary variable because it limits the complexity that can be processed by the communication. A meaning-processing system, for example, can be expected to evaluate its options also in an anticipatory mode because it can entertain and exchange different representations of itself. A knowledge-processing system like a scientific communication system can sometimes reconstruct the system under study--to the extent that one is also able to intervene technologically in what is represented.

**Translation and knowledge generation**

The globalized super-system can be expected to remain latent, but as a next-order selection mechanism it operates with the resilience of a regime. The complex system can be represented only in terms of its "instantiations" or equivalently--along the time dimension--its trajectories. The codes of the communication at the level of these subsystems tend to close the operation by attributing specific meaning to the (Shannon-type) information, but this closure can be altered when exchanges among different codes become historically possible at a next-order level. The changes in meaning due to the exchanges at the level of codes, can be considered "translations" of the information. Translations enable us to change the meaning of an uncertainty from one system of reference to another.

Through translation, discursive knowledge is generated endogenously because one has to distinguish among possible meanings in different contexts. For example, when one discusses energy shortages in political debates, a physicist would first have to translate this concept of "energy" as a shortage of energy *carriers* (e.g., oil or gas) because "energy" is defined in physics as a conserved entity. Just as in the context of physics there can never be a shortage of energy, words often have different meanings in different contexts; yet if one is knowledgeable about the differences between the codifications, translation becomes possible.

The differentiation of the codes possesses an evolutionary function that potentially furthers the development of communication systems. Translations of meaning within the system from one code to another, provide an internal mechanism for the regeneration and transformation of the organization of meaningful expectations. The translation mechanism can therefore be made "functional" to the "reproduction" of the knowledge-base of the social system. Novelty is generated when new representations emerge from the innovative *recombination* of codes.

For example, technological innovation can be considered as the successful generation of new combinations of "demand" and "supply." While "demand" is mainly codified in terms of markets and users, the "supply" side is organized in terms of technologies--and the latter may in turn be based on specific combinations of insights from different specialties. The stabilization of innovation into a new system potentially reorganizes the systems from which it was generated. The emergence of the Internet can be studied in these terms, as can the railway systems in the 19th century (Urry, 2000).

Whereas mutation prevailed as a stochastic mechanism for generating variation in biological systems, it is selective and reflexive interactions among (sub)systems of the social system that generate second-order variation. Each historical stabilization within an evolving system will be continuously disturbed by interactions with other parts of the system. When disturbances can be provided with meaning at various systems levels, the historically embedded codes are expected to change gradually. The recursivity in this process of reflexive refinement (locally) improves the system and generates knowledge endogenously. Meanings which are functional are distinguished from those which are not (or no longer) functional for the reproduction of the communication. Solutions to puzzles can be communicated as potential innovations and then selectively codified in a next round (Kuhn, 1962).

To paraphrase Bateson (1972), this endogenous generation of discursive knowledge can perhaps be defined as the exchange of "meaning which makes a difference." When repeated over time, the different selections generate couplings both horizontally and vertically within the knowledge base of the system (Simon, 1973). Some couplings (and not others) are stabilized and provisionally locked-in (Arthur, 1994) forming communication channels that can be considered as codifications of previous communications. Communication along a provisionally constructed and stabilized channel can be received as signals by specific system(s) whereas higher-order languages, such as scientific discourses enable us to refine the distinctions of signals from noise (Luhmann, 1990).

When a coupling is provisionally stabilized, a next-round of discursive reflection and deconstruction may enable us to reconstruct the system under study and perhaps renew it by searching for a solution to the puzzle different from the ones generated "naturally" or at previous moments in time. In Holland, for example, a polder vegetation can be considered "natural," while the polder as a technical system of water management remains artificial. The social system is increasingly able to replace its historically given base with an evolving knowledge base that operates *analytically* independent on the historical organization of meaning. The discursive reconstruction of codification in relevant exchange processes (e.g., markets, sciences, etc.) enables us to deconstruct an interface in terms of its composing (sub)dynamics and then sometimes reconstruct it more efficiently by shaping technological artefacts. This technological reconstruction of the natural base can be reinforced at the global level and therefore develop beyond the control of individual or group intentions (Marx, 1857).

(Sub)systems shape one another historically along the time axis by providing meaning to their mutual disturbances. The reconstructions increasingly invert the time axis in an evolutionary manner, from within. This process uses another ("incursive") dynamic of the communication. As the interactions gain systemic momentum, the emerging patterns guide the selections with hindsight because the constructs become structured as another network layer. The global perspective, although evolutionarily emerging, may eventually begin to drive the historical development of the system into a self-organizing mode. The "phase space" of alternatives provided by globalization offers possibilities for restructuring the system. This development is knowledge-based and selective because some of the envisioned possibilities are no longer "naturally given."

The globalized system can be hypothesized as selecting upon historical manifestations that are recognized in the present. The representations refer to expectations that can be entertained about other ("adjacent"; cf. Kauffman, 2000) possibilities. Globalization can be considered as an evolutionary process that inverts the time axis by opening and then also organizing a space of envisioned possibilities from which one is able to select incursively (Van Lente and Rip, 1998). The theoretical representations remain constrained by the manifestations because one cannot select upon what has not yet been envisioned.

**Regimes and identities**

Globalization provides an additional dimension to the social system, which can only be accessed by the individual through previously stabilized systems of communication. The individual in an increasingly knowledge-based society has the opportunity to trade-off between stabilization and globalization. At moments when the two operations can be mapped onto each other, an identity can be temporarily maintained. At other moments, one interacts with the social system, and a next-order selection may remain external to the individual. Thus, next-order selection by the social environment and internal reflection at the individual level compete in providing meaning. Both autonomy and inclusion are part of one's personal development in these complex dynamics (Weinstein and Platt, 1969).

The two systems (the individual and the social) can be considered as structurally coupled as they provide complexity to each other (Maturana & Varela, 1980). However, the axes of the systems stand perpendicular to each other, and this difference can be made reflexive within both systems. Whereas the individual is consciously interested in maintaining identity, uncertainty can be expected to prevail in the network mode of communication. The network improves by developing *discursive* knowledge, whereas the individual can feel alienated from ongoing processes of codification in subsystems of social communication in which one is not or only partially included. The networks generate meanings and discursive knowledge at the supra-individual level, whereas the individual seeks to reflect on these meanings from an identifiable perspective.

The social system reflects the complex dynamics of each individual, but in a distributed mode. The interactions among mutual expectations potentially add another dynamic to the aggregate of individual expectations. And as the human carriers continue to communicate in terms of information and meaning, this communication system becomes increasingly shaped so that it enables them to process two or even more dimensions of the communication at the same time without continuously becoming confused. Thus, the development of the communicative capacities of the carriers promotes the further differentiation within, and codification of, the social (sub)systems, and vice versa.

As long as the actors operated on the basis of a single (e.g., cosmologically given) meaning at the level of the social system, this system could still be considered as an identity. As the actors increasingly deconstruct and reconstruct meaning within communication, the difference between the communication of uncertainty and meaning can be reinforced by codification. When this differentiation can further be stabilized--by communicating through symbolically generalized media for example--one can expect the social system to become globalized.

As a historical phenomenon, the globalization of the social system remains part of the social system as one of its subdynamics. The super-system fails to exist physically or biologically, since it is functionally defined as a knowledge-based operation, one of changing what can be observed. It can only be considered as "real" in the sense of critical realism while it exhibits itself in terms of its contingent instantiations (Bhaskar, 1997; Sayer, 2000). The globalized super-system, however, can be considered as another subsystem--one that modulates local communications--in terms of, for example, these communications' symbolic value (Leydesdorff, 1993).

For analytical reasons, a reflecting communication must be distinguished from the communication on which it reflects. Thus, the social system is in need of a medium that enables us to communicate in terms of both information exchanges and in terms of the meaning of this exchange. These two dimensions have not been considered here as different systems, but as differentiations of the linguistic operator: the two layers are not "hard-wired." The distinctions remain reflexively constructed, although such distinctions can be codified historically and stabilized for considerable periods of time.

For example, social meaning can be codified at the supra-individual level such as in the case of the development of a scientific paradigm. The differentiation between common-sense knowledge and scientific (i.e., potentially counter-intuitive) knowledge can be made functional to the further development of the individual and/or the society, but again along different axes. In a scientific discourse, the two dimensions (the embeddedness in a social group as an individual and the cognitive partaking in a scientific development by providing contributions) can no longer be expected to coincide as systems of reference (Gilbert & Mulkay, 1984; Leydesdorff, 1995). Thenceforth, the problems of differentiation and integration increasingly become puzzles for the social coordination and for the individuals living in a knowledge-based society (Bernstein, 1995).

Meaning at the level of the social system is thus different from meaning at the level of the individual. The social level can be considered an outcome of the *interactions* among individuals. This emergence of social meaning has also been called "situational meaning" in symbolic interactionism (Blumer, 1969; Knorr-Cetina & Cicourel, 1981). Human carriers are increasingly able to distinguish reflexively between the social meaning of a communication and their personal meaning. This differentiation can be reinforced as the difference between personal (or tacit) knowledge and discursive (i.e., potentially counter-intuitive) knowledge (Leydesdorff, 2000).

Social order can be entertained as an expectation that is informed by the manifest institutionalizations of the system (Giddens, 1984). Social order, however, consists of expectations being exchanged among individuals realizing their life-cycles in interaction with the contingencies of their biological bodies. Within the life-cycle of an individual certain problems have to be solved: real-life conditions place constraints on the differentiation between possible meanings and the distribution of events (Habermas, 1981, 1987). At the level of a social system, this natural constraint can be relaxed: different solutions can be achieved by various subsystems. The function of differentiation can be expected to remain uncertain, or in flux, more than in the case of a biological system.

As the social system becomes further differentiated into various subsystems, discourses and codes (such as between an economy, scientific discourses, health care systems, etc.)--- the carriers are burdened with making translations among the specific languages. Furthermore exchange media must be developed and potentially globalized in each of these subsystems. Reflexive discourses enable the carriers to make these translations of meanings in various ways. As they are able to communicate about their respective solutions, for example, using public or symbolic media, they may subjectively envisage their historical solution as a global solution. However, with hindsight, each solution at the level of the social system can be recognized as the globalization of a specific medium (e.g., a television channel like CNN). Globalization of the social system can be expected to be differentiated into a range of local solutions that have been provisionally stabilized. The social system continuously loops back into itself (e.g., between local and global) since no informal transgressions are intrinsically forbidden.

A globalization represented at a specific moment in time will be perceived in the future as an instantiation of this subdynamic because the globalization itself remains dynamic. The dynamic perspective contains more complexity than one is able to observe of any discrete instance. This more complex system of communications builds on its subdynamics by selecting from them, but not in any *a priori* prescribed order. In addition to institutions, the observable manifestations include rules and regulations as the dynamic counterparts of the institutions (Giddens, 1979 and 1984). These units of analysis can be considered as the footprints of the communications that have served us hitherto. The organization of communications develops along trajectories that have been institutionalized and codified for historical reasons.

Institutionalization and stabilization are historically observable, whereas codification can be considered as providing meaning to some meanings but not to others. Therefore, the latter operation is part of the incursive process of cultural evolution. As noted, this process inverts the time axis and highlights the construction of meaning within the system as a future-oriented operation that takes place in the present. By providing new meaning to its previous instantiations, the system is increasingly able to rewrite its history; for example, by making new selections from the perspective of hindsight. As the new meanings are increasingly science-based, the rewritings can become increasingly technology-based. The systematic organization of knowledge production (e.g., in R&D laboratories) drives the social system into its globalization.

Here I have followed Luhmann in defining the social system as structurally coupled to human agency. And whereas the social system is generated from and based in the network layer among the carriers at the nodes, the links themselves are not stable. That is; they can be considered as operators that induce change. Virtual communications among machines can then also be constructed using specific media of communication as next-order codifications--for example, into scientific and technical discourses. However, these virtual communications have to be provided with meaning locally, since otherwise they would lose their historical significance.

Thus, the evolutionary metaphor and the historical metaphor are two sides of the same coin in the case of the social system's operation. The co-evolution of incursive and recursive subroutines drives the development of the knowledge base as an emerging dimension of the social system. Geometrical metaphors reduce the algorithmic complexity in the systems under study by providing us with perspectives that can provisionally be stabilized by codification. The codes of the communication span a universe that processes complexity, while the interfaces between the social and the individual ground the communication. This structural coupling sets the discursive stage for what can historically be considered as "natural," "social," or "virtual." The capacities to channel information selectively continuously challenge the (sub)systems and their interactions to explore innovative recombinations of meaning.

**Communication in the knowledge base**

Methodologically, one is able to measure communications in terms of changes in the distributions of what is exchanged. By definition, a redistribution generates a probabilistic entropy. Using the abstract concept of "probabilistic entropy," "variation" and "selection" can be considered as two geometrical metaphors for the study of the same (algorithmic) process: the observable variation provides us with a historical description of the evolutionarily deselected cases. In other words: the concepts of probability and probabilistic entropy provide us with a common denominator for "selection" and "variation" when using the mathematical theory of communication. This reflects a perspective very different from that of (Darwinian) evolution theory in biology, where these two operations have been considered analytically independent. An empirical event can be expected to occur (with a probability); yet the occurrence of events remains uncertain because of selection pressure. At each instance, randomness, or variation, interacts with determination (or selection).

In mathematics, one can deduce a consequence by making an abstract inference, but one can no longer expect to be completely certain (nor completely uncertain) when studying empirical operations. The theoretical specification of the substances of the communication enables us to select the relevant dimensions from an abstract phase space of possible dimensions. A specific theory of communication can consequently be generated in each dimension that can be specified as relevant. The mathematical concepts are thus provided with substantive meaning.

When it is assumed for example, that atoms are exchanged among molecules, chemistry can be developed as a discourse that reflects upon these exchanges (Mason, 1992). When molecules are exchanged, biology can be considered as a discourse that is able to appreciate these processes in theoretical terms (Maturana, 1978). When meaning is communicated, psychology and sociology provide us with relevant reflections (Luhmann, 1986). These latter two discourses refer to different systems of reference, namely individuals and the coordination mechanisms among individuals, respectively. Because of the differentiation within the social system, reflexive study of this system--sociology--can be expected to proliferate its discourses endogenously (Leydesdorff, 1997).

The substantive interpretations of the various subdynamics provide us with metaphors. These metaphors can be thought of as geometrical stabilizations that are useful in the codification of a specific type of specialized communication (Hesse, 1980). They provide analysts with a structure that can sometimes be stabilized (into a scientific discourse) and perhaps even globalized (into a paradigm). As substantively specific, the use of metaphors enables us to handle the algorithmic complexity in the data without becoming continuously confused by the ongoing processes of change at various levels.

Whenever a variation occurs that is not completely random, one selection or another can be hypothesized, since the variation could have been different. Whether the result is perceived as an observable variation or as the result of a (hypothesized) selection depends on the perspective of the analyst. Thus, the concepts of "variation" and "selection" can be considered as two geometrical perspectives on the algorithmic operation of generating probabilistic entropy. The system of reference for the variation is different from that for the selection. In classical evolution theory, for example, variation occurs at the species level, while selection is attributed to nature as a super-system.

The notion of geometrical metaphors used for communication brings language into play (Hesse, 1988). The perspectives can be considered as codified at the supra-individual level, that is, by using one selection or another for the stabilization of the complex dynamics into a discursive (that is, geometrical) representation. This representation can contain an instantiation at one moment in time or entail the development of a trajectory over time. However, an information *calculus* is required for the specification of the probabilistic operation itself (Bar-Hillel, 1955).

The synchronic representation of an instantiation highlights the complexity, whereas the diachronic representation of a trajectory focuses on the dynamics. The complex dynamics, however, develops beyond these representations in terms of fluxes. This global level of communication can be accessed in terms of expectations on the basis of the codified knowledge contained in the representations, for example, with the help of an algorithmic simulation model. In the algorithmic model, the representations are recursively represented as subroutines (Leydesdorff & Dubois, 2003).

**Conclusion**

Human language first enables us to codify the relationship between uncertainty and the meaning of a message. As meaning is generated interactively by using language, it can be considered as a reflexive function of language at the level of the social system (Pask, 1975). I have here deviated from Luhmann's social systems theory by considering language as the "first-order" operating system of society. Languages spontaneously emerge among human beings as "natural languages" (Pinker, 1994); however languages can also be cultivated using higher-order codifications, for example, in scientific discourses.

Luhmann in 1984 proposed that "meaning" was itself the operator of social systems. "Meaning" was therefore considered as a kind of transcendental precondition of both reflection and language, but not as a consequence of the linguistic operation itself (Luhmann, 1971). Luhmann had not yet really reflected upon the idea of meaning from a more abstract perspective such as the mathematical theory of communication, or non-linear dynamics. Luhmann's focus on using historical examples implied that the globalization of meaning was considered a virtual operation that could only be discussed with reference to Husserl's transcendental phenomenology.

The social construction of meaning was analyzed as thoroughly contingent, by scholars working in the pragmatist tradition and in symbolic interactionism (Blumer, 1969; Mead, 1934). The exchange of meaning provides communication with a second layer that reflects and interacts with the first layer of uncertain exchanges in the social system. I propose that language can be considered as the evolutionary achievement that facilitates communication by using these two channels simultaneously. This duality enables us to capture the complexity of the algorithmic operation, a problem having two sides: an *a priori* expectation and an *a posteriori* update value. Meaning draws in the present upon the information exchanged when it "incurs" on the information. It is also continuously generated as a recursive (i.e., historical) update of previous meaning(s). The difference between these two operations can be observed in social systems since meaning can be codified at different levels of the same system.

The organization of scientific knowledge production for example, has led to the differentiation between the context of justification and the context of discovery (Popper, 1935). Knowledge produced in the context of discovery can be recognized as a knowledge claim when submitted to a journal for further communication. Accordingly, the two contexts provide different meanings to the same knowledge and distinctions in the epistemological status of this knowledge (Gilbert & Mulkay, 1984).

The meaning of a communication can be changed with hindsight, by being exchanged. It can temporarily be codified and historically stabilized, as for example, in social institutions. The incursive subroutine, however, doubles the historical authenticity of the social system by providing it with an evolutionary layer that is future-oriented in the present. Thus the complex social system as the subject of cultural evolution is able to reconstruct its history continuously, and from different perspectives, while operating reflexively and in a distributed mode. Because of this distributed mode, the reconstruction cannot be completed without a further selection. Through translation by a receiving discourse, a previous codification can sometimes be stabilized since the meaning may have been rewritten as historically specific.

The appreciation of incursion *within* the recursive routines enables us to reconstruct the historical dimension of these processes. In other words, the system re-doubles in the representation because the linguistic operation itself is layered. A knowledge-based system can entertain more than two representations because what was first considered as levels can also be considered as dimensions of codification along different axes. Over time--yet another dimension--the operation is neither completely historical nor completely reconstructive: it contains uncertainty also in the time dimension. Cultural evolution, once unleashed as an interaction among these degrees of freedom, drives the communicative competencies of the carriers at an increasing speed since the social system provides us with ranges of options for realization at each reflexive turn.

While reflection is often associated with a turn of 180 degrees, the addition of another dimension to the social system assumes a turn of only 90 degrees. Interaction terms among these perpendicular dynamics of reflection can lead to non-linear recombinations of meaning. However, the number of possible combinations--and therefore the selection pressure--increases exponentially with the number of relevant dimensions. As the awareness of these dimensions is absorbed, the carriers can experience this pressure as globalization of the social system. Insofar as the dimensions can be specified as hypotheses, they can be made the subject of discursive reasoning. The analytical specification of these systems of reference can be expected to become crucial for the further development of knowledge-based communication systems.

- The author is grateful to Gretchen M. Dee and Iina Hellsten for detailed comments on a previous draft.

Loet Leydesdorff is Senior Lecturer for "Communication and Innovation in Science and Technology" at the Amsterdam School of Communications Research (ASCoR) of the University of Amsterdam. See at http://www.leydesdorff.net/list.htm for his other publications.